\documentclass[twocolumn]{aastex62}

\usepackage{graphicx}
\usepackage{xcolor}
\usepackage{natbib}
\usepackage{amsmath}
\usepackage[subfigure]{tocloft}
\usepackage{subfigure}
\usepackage{CJK}

\newcounter{RomanNumber}

\newcommand{\ue}{\mathrm{e}}

\received{xxx}
\revised{xxx}
\accepted{xxx}

\graphicspath{{./}{fig/}}

\shorttitle{PTDE}
\shortauthors{Chen \& Shen}

\begin{document}
\begin{CJK*}{UTF8}{gbsn}

\title{Light Curves of Partial Tidal Disruption Events}

\author{Jin-Hong Chen (陈劲鸿)}
\affiliation{School of Physics and Astronomy, Sun Yat-Sen University, Zhuhai, 519000, P. R. China}
\author{Rong-Feng Shen (申荣锋)}
\affiliation{School of Physics and Astronomy, Sun Yat-Sen University, Zhuhai, 519000, P. R. China}

\email{chenjh258@mail2.sysu.edu.cn, shenrf3@mail.sysu.edu.cn}

\begin{abstract}
Tidal disruption events (TDEs) can uncover the quiescent black holes (BHs) at the center of galaxies and also offer a promising method to study them. In a partial TDE (PTDE), the BH's tidal force cannot fully disrupt the star, so the stellar core survives and only a varied portion of the stellar mass is bound to the BH and feeds it. We calculate the event rate of PTDEs and full TDEs (FTDEs). In general, the event rate of PTDEs is higher than that of FTDEs, especially for the larger BHs. And the detection rate of PTDEs is about dozens per year by Zwicky Transient Factory (ZTF). During the circularization process of the debris stream in PTDEs, no outflow can be launched due to the efficient radiative diffusion. The circularized debris ring then experiences viscous evolution and forms an accretion disk. We calculate the light curves of PTDEs contributed by these two processes, along with their radiation temperature evolution. The light curves have double peaks and the spectra peak in UV. Without obscuration or reprocessing of the radiation by an outflow, PTDEs provide a clean environment to study the circularization and transient disk formation in TDEs.
\end{abstract}

\keywords{accretion, accretion disks - black hole physics - galaxies: nuclei }

\section{Introduction}
\label{introduction}
In the nucleus of a galaxy, when an unlucky star is occasionally perturbed into an orbit on which it comes too close to the central supermassive black hole (SMBH), it will be destroyed by the tidal force \citep{Rees_Tidal_1988}. In such a TDE, the accretion of the debris produces a flare and illuminates the galaxy for a period of months to years. A large sample of TDEs can uncover the hidden population of SMBHs in the center of quiescent galaxies, and provide a promising method to measure the properties of BHs and to study the physics of accretion.

The main observational properties of dozens of discovered candidate TDEs are: bright in UV/optical band with almost constant temperature about $2-4\times 10^4$ K near the peak of the luminosity, and some of them show X-rays which might come from the accretion disk. The photosphere radius of the UV/optical radiation inferred by assuming a blackbody spectrum is $\sim 10^{15}$ cm, which is larger than the circularization radius $R_{\rm c} \sim 10^{13}$ cm \citep{Gezari_An_2012,Holoien_ASASSN14ae_2014,Holoien_Six_2016}.

It is commonly considered that the UV/optical emission originates from the self-collision near the apocenter \citep{Piran_Disk_2015} or comes from the reprocessing layer, which is produced in the circularization process as the stretched stream shocks itself near the apocenter due to the apsidal precession \citep{Jiang_Prompt_2016,Lu_Self_2020}, or is driven by the super-Eddington accretion process in the accretion disk \citep{Strubbe_Optical_2009, Lodato_Multiband_2011,Metzger_A_2016}.  Though these models can explain many TDE candidates, the issues including the circularization process and the formation of the accretion disk are still unclear.

If the pericenter of the tidally disrupted star is slightly farther away than the tidal radius, it cannot be fully destroyed and therefore retains the core after the encounter. \cite{Guillochon_Hydrodynamical_2013} found the critical value of the so called penetration factor that separates TDEs into full TDEs (FTDEs) and partial TDEs (PTDEs) by hydrodynamical simulations.

In this paper, we consider some PTDEs which will not produce outflow (wind) driven by the super-Eddington accretion or the self-collision. Without the outflow, these PTDEs provide a clean environment to study the circularization process of the debris stream and how the accretion disk forms. 

In Section \ref{typical}, we describe the characteristic dynamical properties of PTDEs. In Section \ref{cir_process} - \ref{temperature_evolution}, we calculate the light curve and the temperature of PTDEs in the circularization process and in the viscous evolution. In Section \ref{dependence}, we study the dependence on the mass of the BH and the disrupted star. In section \ref{event_detection_rate}, we estimate the ratio of the event rate between PTDEs and FTDEs, and calculate the detection rate of PTDEs. We summarize and discuss the results in Section \ref{conclusion}.

\section{Characteristic dynamical properties}
\label{typical}
When a star approaches the tidal radius $R_{\rm T} = R_*(M_{\rm h}/M_*)^{1/3}$ \citep{Rees_Tidal_1988,Phinney_Manifestations_1989} in a parabolic orbit, it can be disrupted by the tidal force from an SMBH. Here $M_{\rm h} \equiv {M_6} \times 10^6\ \rm{M_{\odot}}$, $R_* \equiv r_* \times R_{\odot}$, $M_* \equiv m_* \times \rm{M_{\odot}}$ are the BH's mass, star's radius and mass, respectively. Some materials within the star during the encounter can overcome the self-gravitional force and become unbound to the star, but others are not and left a core after the encounter. We can use the penetration factor $\beta \equiv R_{\rm T} / R_{\rm p}$ to quantify this effect. Here $R_{\rm p}$ is the pericenter radius, in unit of the BH's Schwarzschild radius $R_{\rm S} = 2GM_{\rm h}/c^2$, it is
\begin{equation}
R_{\rm p} \simeq 23\ \beta^{-1} M_6^{-2/3} r_* m_*^{-1/3}\ R_{\rm S}.
\end{equation}

The hydrodynamic simulation results of \citet[hereafter G13]{Guillochon_Hydrodynamical_2013} showed that a star is fully disrupted when $\beta \ge \beta_{\rm d}$, and a partial disruption happens when $\beta < \beta_{\rm d}$. For stars with the polytropic index $\gamma = 4/3$, $\beta_{\rm d}=1.85$, and for $\gamma = 5/3$, $\beta_{\rm d}=0.9$. Notice that there are some other works indicating similar results \citep[e.g.,][$\beta_{\rm d}=0.92$ for $\gamma = 5/3$ and $\beta_{\rm d}=2.01$ for $\gamma = 4/3$]{Mainetti_The_2017}. \cite{Ryu_Tidal1_2020} found different results for different stellar mass by using stellar evolution code MESA.

When the star is tidally disrupted by an SMBH, the debris would have a range in specific energy due to their locations in the SMBH's potential well. In the "frozen-in" model \citep{Lodato_Stellar_2009}, the energy spread is $\epsilon = \pm GM_{\rm h}x/R_{\rm p}^2$, where $x$ is the distance from the center of the star. The most bound debris with a specific energy\footnote{Notice that this is different from the case of fully disruption ($\beta \gtrsim 1$), whose specific energy of the most bound material is $\epsilon_0 \simeq GM_{\rm h} R_*/R_{\rm T}^2$, because in the latter case this energy is determined at $R_{\rm T}$, not at $R_{\rm p}$ \citep{Guillochon_Hydrodynamical_2013}.} of 
\begin{equation}		\label{eq:eps0}
\epsilon_0 \simeq \frac{GM_{\rm h} R_*}{R_{\rm p}^2} \simeq \frac{GM_{\rm h}}{2a_0}
\end{equation}
is the first to return to the pericenter, where 
\begin{equation}
a_0 \simeq \frac{R_{\rm p}^2}{2R_*}
\label{semimajor}
\end{equation}
is the semi-major axis of its orbit, corresponding to an eccentricity of $e_0 = 1 - R_{\rm p}/a_0$.
The period of this orbit 
\begin{equation}
t_{\rm fb} = 2 \pi \sqrt{a_0^3/GM_{\rm h}} \simeq 41\ \beta^{-3} M_6^{1/2} r_*^{3/2} m_*^{-1}\ {\rm day}
\label{pmb}
\end{equation}
determines the characteristic time-scale of the debris fallback.

Typically, less bound debris follows the most bound debris to return, at a rate that decays as $t^{-5/3}$ based on the constant $dM/dE$ \citep{Rees_Tidal_1988, Phinney_Manifestations_1989,Ramirez-Ruiz_THE_2009}. Actually the fallback rate is more complex, even the tail of the fallback rate in PTDE is much steeper than the power law $-5/3$ \citep{Guillochon_Hydrodynamical_2013, Ryu_Tidal3_2020,Coughlin_Partial_2019}. 

In order to get more flexible results and to obtain the total bound mass to fall back, we adopt the fitting formulae of the simulation results in G13 for the total fallback mass $\Delta M$. The stellar polytropic index is $\gamma=5/3$ in this paper. \cite{Ryu_Tidal3_2020,Law_Stellar_2020} obtained similar results from the simulations of the disruption, considering the more realistic stellar structure output by the stellar evolution code MESA. Here since we only propose to predict the general features of PTDEs, we omit the study of the dependence on the stellar evolution.

In this paper, we only consider the bound debris left by the encounter, which will be accreted by the SMBH afterward. The remnant core could possibly be bound to the SMBH after the encounter, and fall back to be disrupted again \citep{Ryu_Tidal3_2020}. However, the orbital periods of the remnant core are $\simeq 400 - 40,000$ yr, therefore too long for detection. Furthermore, whether the remnant core is bound or unbound is still unclear \citep{Manukian_Turbovelocity_2013,Ryu_Tidal3_2020}.

\section{Circularization}		\label{cir_process}

\subsection{Stream Crossing}			\label{sc_cir}

\begin{figure}		
\centering
 \includegraphics[scale=0.5]{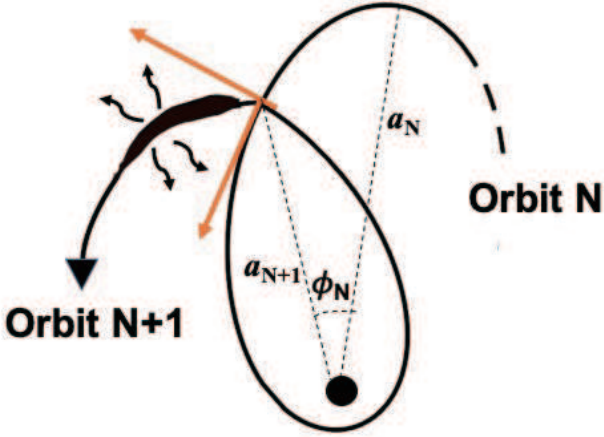}
\caption{A sketch of the stream crossing. After the stream precesses by an angle $\phi_{\rm N}$ on the $N$-th orbit, it collides with itself, thus it losses energy and moves to $N+1$-th orbit. Because of the efficient radiative diffusion in PTDEs, most of the thermal energy produced by the collision radiates from a small region near the collision point. The semi-major axis of orbits $N$ and $N+1$ are $a_{\rm N}$ and $a_{\rm N+1}$, respectively. The velocity of the two colliding components are shown as the orange arrows. This sketch is adapted from \cite{Bonnerot_Long_2017}.}
\label{sketch}
\end{figure}

Many studies indicate that the circularization of the debris is  possible due to the general relativistic apsidal precession, which is crucial for the formation of accretion disk \citep{Rees_Tidal_1988,Hayasaki_Finite_2013,Dai_Soft_2015,Bonnerot_Disc_2015,Bonnerot_Long_2017}. Upon each passage of the pericenter, the stream precesses by a small angle $\phi \sim R_{\rm S} / R_{\rm p}$, thus undergoes a succession of self crossings, which dissipates an amount of the specific energy. Consequently, the apocenter of the stream's new orbit moves closer to the BH, and its eccentricity decreases. The stream crossing is illustrated in Figure \ref{sketch}.

In the $N$-th orbit, the specific energy $\epsilon_{\rm N}$, semi-major axis $a_{\rm N}$, and specific angular momentum $j_{\rm N}$ are related as in: $\epsilon_{\rm N} = GM_{\rm h}/(2 a_{\rm N})$, and $j_{\rm N}^2= GM_{\rm h} a_{\rm N}(1- e_{\rm N}^2)$. Under the assumption of completely inelastic collision, the dissipated energy per orbit is (\cite{Bonnerot_Long_2017}, also see \cite{Dai_Soft_2015})
\begin{equation}		\label{eq:delta-eps}
\Delta \epsilon_{\rm N} \simeq \frac{9}{2} \pi^2 \frac{e_{\rm N}^2}{c^4} \left(\frac{GM_{\rm h}}{j_{\rm 0}}\right)^6 = \Delta \epsilon_0 \frac{e_{\rm N}^2}{e_0^2}
\end{equation} 
assuming the stream's angular momentum is conserved during the crossings, i.e., $j_{\rm N} = j_0$.
The dissipated energy during the first crossing is
\begin{equation}		\label{eq:delta-eps0}
\Delta \epsilon_0 = \frac{9}{16} \frac{\pi^2  e_0^2}{(1+e_0)^3} \left(\frac{R_{\rm S}}{R_{\rm p}}\right)^3 c^2
\end{equation} 

When the stream is eventually fully circularized, i.e., $e_N=0$, then $\epsilon_{\rm N}$ shall take its final value $\epsilon_{\rm c} = GM_{\rm h} / (2R_{\rm c})$, where
\begin{equation}
R_{\rm c}= R_{\rm p}(1+e_0)		\label{eq:Rc}
\end{equation}
is the so-called circularization radius.

\subsection{Energy Dissipation Rate History}		\label{dissipation_cir}
We assume that in the circularization phase the energy dissipation mainly comes from the self-collision of the stream. The viscous dissipation is relatively weak during this phase, but it will become important after the circularization (see the discussion in section \ref{conclusion}). Furthermore, we will consider only the part of the stream that is made of the most bound debris, i.e., the ``main stream'', since it returns earlier and comprises the major portion of the total stream mass, and neglect the tail of the stream. Due to apsidal precession, this stream crosses and collides with itself in successive orbits, reducing its energy. 

Adopting the iterative method in \cite{Bonnerot_Long_2017}, the specific energy dissipation rate is $\Delta \epsilon_{\rm N}/t_{\rm s, N}$ in the $N$-th orbit, such that the dissipation rate can be estimated by $\Delta M \Delta \epsilon_{\rm N}/t_{\rm s, N}$ during the circularization, where $t_{\rm s, N}$ is the orbital period of $N$-th orbit. Assuming efficient radiative cooling, the luminosity is equal to the dissipation rate, i.e.,
\begin{equation}
\label{eq:Lc}
L_{\rm N} \simeq \Delta M \frac{\Delta \epsilon_{\rm N}}{t_{\rm s, N}}.
\end{equation}
Here we assume the mass of ''main stream'' equals to the total fallback mass and neglect the tail of returning stream. The reason is that the main stream contains most of the fallback mass after the first collision, which is approximated to be $\Delta M (1-(4t_{\rm fb}/t_{\rm fb})^{-5/4}) \simeq 0.82\ \Delta M$ by assuming $\dot M_{\rm fb} \propto (t/t_{\rm fb})^{-9/4}$ \citep{Coughlin_Partial_2019,Miles_Fallback_2020}.

The mass of the main stream is very sensitive to $\beta$. When $\beta > 0.5$ the tidal disruption occurs \citep{Ryu_Tidal1_2020}. In this paper, we calculate three cases of PTDEs with $\beta = 0.55, 0.6, 0.7$, whose total fallback mass are $\Delta M \simeq 0.0048, 0.0254, 0.1222\ M_*$, respectively \citep{Guillochon_Hydrodynamical_2013}.

Alternatively, we can write the energy dissipation rate history in a differential form: $\dot \epsilon(t) = d \epsilon/dt= \Delta \epsilon_{\rm N}/t_{\rm s, N}$. Substituting the orbital period-energy relation $t_{\rm s, N} \equiv 2\pi GM_{\rm h} /(2 \epsilon_{\rm N})^{3/2}$ and Equation (\ref{eq:delta-eps}), we get a differential equation of $\epsilon$:
\begin{equation}		\label{eq:diss}
\dot \epsilon = \frac{\Delta \epsilon_0}{t_{\rm fb}} \frac{1}{e_0^2} \left(1 - \frac{\epsilon}{\epsilon_{\rm c}}\right) \left(\frac{\epsilon}{\epsilon_0}\right)^{3/2}.
\end{equation}
When $a_{\rm N} \gg R_{\rm c}$, $\epsilon /\epsilon_{\rm c} \ll 1$, the factor $1-\epsilon / \epsilon_{\rm c}$ can be dropped, then one can determine the time-scale of circularization by solving for $\epsilon(t)$ from the above equation. Letting $e_0 \sim 1$, we obtain the circularization time-scale
\begin{equation}		\label{circularization}
\begin{split}
t_{\rm cir} &\simeq 2\frac{ \epsilon_0}{\Delta \epsilon_0} t_{\rm fb}
\\
&\simeq 8 \ \beta^{-1} M_6^{-5/3} m_*^{-1/3} r_*^2 \ t_{\rm fb},
\end{split}
\end{equation}
for PTDEs. The same formula was obtained in \cite{Bonnerot_Long_2017} along a different approach.

For the PTDE with $\beta = 0.5$, it needs to spend a duration $\sim 16\ t_{\rm fb}$ to form a circular disk. Other mechanism, e.g. the magneto-rotational instability (MRI), may cause momentum exchange and speed up the circularization process \citep{Bonnerot_Long_2017, Chan_Magnetorotational_2018}. We assume the viscous effects are weak in the circularization stage. We will further explore this issue in section \ref{conclusion}.

\begin{figure}		
\centering
 \includegraphics[scale=0.5]{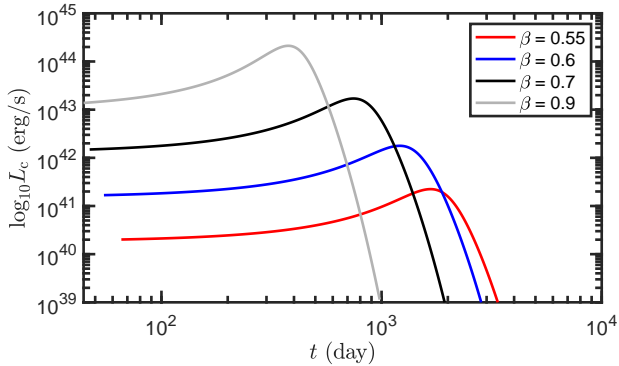}
\caption{Bolometric luminosity history during the circularization process for the disruption of a star ($m_*=r_*=1$) by a $10^6\ \rm{M_{\odot}}$ SMBH. The grey lines represent the borderline FTDEs which have $\beta=\beta_{\rm d}=0.9$, and others belong to PTDEs. It is calculated by Equation (\ref{eq:diss}).}
\label{Lc_m6}
\end{figure}

Furthermore, from Equation (\ref{eq:diss}) it is straightforward to find that the peak dissipative luminosity per mass is
\begin{equation}		\label{eq:dissp}
\dot \epsilon_{\rm p} = \frac 25 \left(\frac35\right)^{3/2}\frac{1}{e_0^2}\left(\frac{\epsilon_{\rm c}}{\epsilon_0}\right)^{3/2}\frac{\Delta \epsilon_0}{t_{\rm fb}}.
\end{equation}
Then using Equation (\ref{eq:eps0}) and (\ref{eq:delta-eps0}), we get the peak luminosity
\begin{equation}		\label{eq:Lcp}
L_{\rm p} = 6 \times 10^{42} \left(\frac{\Delta M}{0.01\ M_{\odot}}\right)\beta^{9/2} M_6^2 m_*^{3/2} r_*^{-9/2}\ {\rm erg\ s^{-1}}.
\end{equation}

Using the differential form, i.e., Equation (\ref{eq:diss}), we can rewrite the circularization luminosity as $L_{\rm c}(t) \simeq \Delta M \dot \epsilon(t)$ and plot it in Figure \ref{Lc_m6}. The result of this differential form is equivalent to that of iterative form (Equation (\ref{eq:Lc})), and it provides a clear relation between the parameters, hence we adopt it in the following calculations.

\subsection{Photon Diffusion}		\label{photon_diffusion}
The luminosity estimate above assumes that photons can diffuse efficiently. However, we should examine the issue of diffusion efficiency more carefully. After the self-collision, the gas is heated, the photon needs some time to diffuse out. The diffusion time-scale determines the observed luminosity and the spectrum. If the diffusion time-scale is much shorter than the orbital period, the radiation mainly emerges from the stream near the collision position. Otherwise, the thermal energy will be accumulated during the circularization process. 

The radiative diffusion time-scale after the shock is given by $t_{\rm diff} \simeq \tau h_{\rm s}/c$, where $c$ and $h_{\rm s}$ are the light speed and the height of the stream, respectively. And the optical depth of the stream after the shock is $\tau \simeq \kappa_{\rm es} \rho h_{\rm s}$, where $\kappa_{\rm es}\simeq 0.34\ {\rm cm^2\ g^{-1}}$ is the opacity for electron scattering for a typical stellar composition \footnote{The atoms are ionized after the shock, and the temperature is high, so that the electron scattering dominates the opacity.}.

Assuming the stream is homogeneous in the interior, the density of the stream after the shock is estimated by
\begin{equation}		\label{eq:density}
\rho = \frac{\Delta M}{4 \pi a_{\rm s} h_{\rm s} w_{\rm s} }.
\end{equation}
Here $w_{\rm s}$ is the width of the stream. And the perimeter of the orbit is $\sim 4 a_{\rm s}$, where $a_{\rm s}$ is the semi-major axis of the orbit. Then the diffusion time-scale after the shock can be written as
\begin{equation}		\label{eq:tdiff}
\begin{split}
t_{\rm diff} &\simeq \kappa_{\rm es} \frac{\Delta M}{4\pi c a_{\rm s}}\left(\frac{h_{\rm s}}{w_{\rm s}}\right)
\\
&\simeq 1.4 \times 10^{-2}\ \left(\frac{\Delta M}{0.01\ \rm{M_{\odot}}}\right) \left(\frac{h_{\rm s}}{w_{\rm s}}\right) \times
\\
&\left(\frac{a_{\rm s}}{a_{\rm 0}}\right)^{-5/2} \beta^5 M_6^{-7/6} r_*^{-5/2} m_*^{5/3}\ t_{\rm s},
\end{split}
\end{equation}
where $t_{\rm s}$ is the orbital period of the stream. In order to estimate the evolution of diffusion time-scale during the circularization process, we need to consider the evolution of the apocenter radius and the height-to-width ratio of the stream.

The apocenter radius becomes smaller as the circularization process carries on. The change of height-to-width is complicated, since it evolves under the gravity and pressure force. In \cite{Bonnerot_Long_2017}, they assume $h_{\rm s}/w_{\rm s} = 1$ during the circularization process. We should relax this assumption here, because the SMBH's gravity will restrict the expansion of stream in the vertical direction, and the width of stream increases slightly due the viscous shear. Therefore, when the stream is cold at later times, the stream will become geometrically thin \citep{Bonnerot_Disc_2015}, so the height-to-width ratio should be $h_{\rm s}/w_{\rm s} \ll 1$.

We set $h_{\rm s}/w_{\rm s} \simeq 1$ at the beginning of the circularization process, and let $h_{\rm s}/w_{\rm s} \simeq 10^{-2}$ when the circularization process completes, which is consistent with the geometrically thin ring (or disk) \citep{Kato_Black_1998}. To account for a smooth transition we adopt the following form for the evolution of the height-to-width ratio
\begin{equation}
\label{eq:htow}
\frac{h_{\rm s}}{w_{\rm s}} \simeq 10^{-2\frac{t}{t_{\rm cir}}}.
\end{equation}

\begin{figure}
\centering
 \includegraphics[scale=0.5]{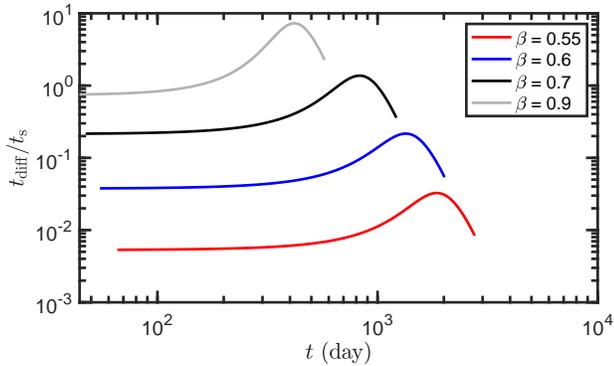}
\caption{Ratio of the radiative diffusion time-scale and the period of the orbit for the disruption of a star ($m_*=r_*=1$) by a $10^6\ \rm{M_{\odot}}$ SMBH. The colors represent different penetration factor $\beta = R_{\rm T} / R_{\rm p}$. It is calculated by plugging Equation (\ref{eq:htow}) into Equation (\ref{eq:tdiff}).}
\label{diffusion}
\end{figure}

We show the history of $t_{\rm diff}/t_{\rm s}$ in Figure \ref{diffusion}. It shows that the radiative diffusion is efficient during the whole circularization process for $\beta \sim 0.5$, but not for $\beta \sim 0.9$. 

If the radiative diffusion is efficient, i.e. $t_{\rm diff} \lesssim t_{\rm s}$, then the thermal energy will not be accumulated in each orbit. At the early time when $a_{\rm s} \sim a_{\rm 0}$, the photon can diffuse out efficiently before the next shock for PTDEs. However, for those FTDEs with $\beta \gtrsim \beta_{\rm d}$, the bound mass $\Delta M \sim 0.5\ M_\odot$, the diffusion time-scale $t_{\rm diff} \gtrsim t_{\rm s}$ and thus the photon cannot diffuse efficiently. 

Since we focus on the PTDEs, the assumption of efficient radiative cooling is reasonable. Therefore we do not consider the photon diffusion in the calculations of the luminosity of circularization hereafter.

\section{The disk viscous evolution}
\label{viscous_evolution}
After the circularization process, the stream settles into the radius $R_{\rm c}$ with the width $w_0$, and the viscous evolution becomes important. In this section we review the basic disk equations, then study the subsequent evolution by an analytic calculation. Then using a numerical model of viscous evolution, we test the analytical results and obtain the detailed spectral evolution.

\subsection{Disk Equations}
The energy balance during the viscous evolution is given by $Q^+=Q^-_{\rm adv}+Q^-_{\rm rad}$. Here the viscous heating rate per unit surface area is 
\begin{equation}
Q^+=\frac94 \nu\Sigma\Omega^2.
\label{Q+}
\end{equation}
The radiative cooling rate is 
\begin{equation}
Q^-_{\rm rad}=\frac{4acT_{\rm c}^4}{3\kappa \Sigma},
\label{Qrad}
\end{equation}
and the advective term is given by
\begin{equation}
Q^-_{\rm adv} = \frac{\dot M_{\rm acc}}{2\pi R^2}\frac{P_{\rm tot}}{\rho} \xi,
\label{Qadv}
\end{equation}
where $\dot M_{\rm acc} = 3 \pi \nu \Sigma$ is the local accretion rate and $\xi$ is close to unity \citep{Frank_Accretion_1985}. Here $\Sigma$, $\Omega$, $a$, and $T_{\rm c}$ are the local surface density, the local angular velocity, radiation constant, and the mid-plane temperature, respectively. 

The total pressure is $P_{\rm tot} = P_{\rm rad} + P_{\rm gas} = aT_{\rm c}^4/3+\rho k_{\rm b} T_{\rm c}/(\mu m_{\rm p})$. Here $k_{\rm b}$, $\mu = 0.6$, and $m_{\rm p}$ are the Boltzmann constant, mean particle weight, and proton mass, respectively. The local density is $\rho \simeq \Sigma/(2H)$. Here the height-scale is given by the hydrostatic equilibrium, i.e.,
\begin{equation}
H=\Omega^{-1} \left(\frac{P_{\rm tot}}{\rho}\right)^{1/2}.
\end{equation}

We adopt the $\alpha_{\rm g}$-viscosity \citep{Sakimoto_Accretion_1981}, i.e.,
\begin{equation}
\nu=\frac{2\alpha P_{\rm gas}}{3\Omega \rho}
\label{viscosity}
\end{equation}
to calculate the viscous evolution.

The local opacity is $\kappa=\kappa_{\rm es} + \kappa_{\rm R}$, where the electron scattering opacity is dominated by Thompson electron scattering, i.e., $\kappa_{\rm es} = 0.2(1+X)\ {\rm cm^2g^{-1}}$, and the Kramer's opacity is given by $\kappa_{\rm R} = 4 \times 10^{25} Z(1+X) \rho T^{-3.5}\ {\rm cm^2g^{-1}}$. The gas composition we adopt in this paper is the solar composition, i.e., $X=0.71$, $Y=0.27$, and $Z=0.02$.

\subsection{Analytical Calculation}
\label{analytic}
The viscous evolution was considered by \cite{Cannizzo_The_1990}, and we summarize it here. We assume the disk is in thermal equilibrium between viscous heating and radiative cooling, i.e., $Q^+ = Q^-_{\rm rad}$. Using these relations we obtain the temperature
\begin{equation}
T_{\rm c}=\left( \frac98 \frac{\alpha \kappa}{ac} \Omega \Sigma^2 \frac{k_{\rm b}}{\mu m_{\rm p}} \right)^{1/3}.
\label{Tc}
\end{equation}

The viscous time-scale is given by 
\begin{equation}
t_{\nu} = R^2/\nu.
\label{viscous_time}
\end{equation}
It is the function of time and radius, but we can average it with respect to $R$ by letting $R=R_{\rm d}$ and $\Sigma=\Delta M/(2\pi R_{\rm d}^2)$, where $R_{\rm d}$ is the average radius of the disk (ring). Substituting the average surface density and radius into Equation (\ref{Tc}), and using Equations (\ref{viscosity}) and (\ref{viscous_time}), we obtain
 \begin{equation}
 \begin{split}
 &t_{\nu} = C R_{\rm d}^{7/3} M_{\rm d}^{-2/3},
 \\
 &C \equiv \alpha^{-4/3} \left(\frac{k_{\rm b}}{\mu m_{\rm p}} \right)^{-4/3}\left(\frac{3acGM_{\rm h}}{\kappa} \right)^{1/3},
 \end{split}
 \label{tv}
 \end{equation}
 where $M_{\rm d}$ is the total mass of the disk (ring). 

At the beginning $R_{\rm d}=R_{\rm c}$ and $M_{\rm d}=\Delta M$, the initial viscous time-scale is
\begin{equation}
\begin{split}
t_0& = C R_{\rm c}^{7/3} (\Delta M)^{-2/3}
\\
& = 13.4\ \alpha^{-4/3} \kappa^{-1/3} \beta^{-7/3} M_6^{10/9} m_*^{-7/9} r_*^{7/3} \times
\\
&\left(\frac{\Delta M}{0.01\ \rm{M_{\odot}}} \right)^{-2/3}\ {\rm yr},
\end{split}
\label{t0}
\end{equation}
which is the time-scale of the ring-to-disk phase. Notice that the ring-to-disk time-scale does not depend on the initial width of the ring.

In order to estimate the bolometric luminosity, we assume the accretion rate is
\begin{equation}
\frac{dM_{\rm d}}{dt} = -\frac{M_{\rm d}}{t_{\nu}},
\label{macc}
\end{equation}
and keep the total angular momentum $J_{\rm d} = M_{\rm d} (GM_{\rm h}R_{\rm d})^{1/2}$ constant during the ring-to-disk and disk phase \citep{Kumar_Mass_2008}, i.e.,
\begin{equation}
M_{\rm d}^2 R_{\rm d} = (\Delta M)^2 R_{\rm c}.
\end{equation}
Then Equation (\ref{tv}) becomes $t_{\nu}=t_0(M_{\rm d}/\Delta M)^{-16/3}$. Substituting it into Equation (\ref{macc}) and assuming a constant opacity, one obtains the accretion rate
\begin{equation}
\dot M_{\rm acc} = \frac{\Delta M}{t_0} \left(1+\frac{16}{3}\frac{t}{t_0} \right)^{-19/16}.
\label{accretion_disk}
\end{equation}
The bolometric luminosity can be written as $L_{\rm disk} = \eta \dot M_{\rm acc} c^2$. The efficiency is $\eta = 1/12$ for Schwarzschild BH. We plot it in Figure \ref{Lbol} with the electron scattering assumption $\kappa=\kappa_{\rm es}$ to compare with the numerical results. According to the results in Figure \ref{Lbol}, we can estimate the peak luminosity in viscous evolution by
\begin{equation}
\begin{split}
L_{\rm disk, p} &\simeq \eta c^2 \dot M_{\rm acc}(t_0) 
\\
&\simeq 2 \times 10^{41}\ \alpha^{4/3} \left(\frac{\Delta M}{0.01\ \rm{M_{\odot}}} \right)^{5/3} \times 
\\
&\beta^{7/3} M_6^{-10/9} m_*^{7/9} r_*^{-7/3}\ {\rm erg\ s^{-1}}.
\label{eq:Lbc}
\end{split}
\end{equation}
When $t\gtrsim t_0$, the luminosity is $\propto t^{-1.2}$, which is same as the self-similar result in \cite{Cannizzo_The_1990}. Notice that the early part ($t < t_0$) of the solution is a rough approximation, because at this stage the accretion rate in Equation (\ref{accretion_disk}) is not necessarily the mass inflow rate at the inner boundary of the disk (i.e., near the BH's event horizon) due to a likely viscous diffusion delay. A more rigorous way to explore this early phase is described below.

\subsection{Numerical Calculation}
\label{numerical}
The viscous evolution is governed by the diffusion equation of the surface density \citep{Frank_Accretion_1985}, i.e.,
\begin{equation}
\frac{\partial \Sigma}{\partial t} = \frac{3}{R}\frac{\partial}{\partial R} \left[R^{1/2} \frac{\partial}{\partial R} (\nu \Sigma R^{1/2}) \right].
\end{equation}

We list the assumptions and initial conditions of the numerical model here:
\begin{itemize}
\item We assume the ring is axisymmetric with a Gaussian surface density profile centered at the circularization radius $R_{\rm c}$
\begin{equation}
\Sigma (R)=\zeta \frac{\Delta M}{R_{\rm c} w_0} {\rm exp} \left[-\left(\frac{R-R_{\rm c}}{2 w_0}\right)^2\right],
\end{equation}
where $\zeta \simeq 1/(4 \pi^{3/2})$ is a coefficient that satisfies
\begin{equation}
\int^{\infty}_{R_{\rm in}} \Sigma (R) 2 \pi R\ dR = \Delta M.
\end{equation}
We set the initial width of the ring as $w_0 = 0.1 R_{\rm c}$.

\item We set the inner boundary condition to be $R_{\rm in}=R_{\rm ISCO}=3R_{\rm S}$ corresponding to a Schwarzschild BH. Once the matter arrives the boundary, it is removed.

\item For simplicity we adopt the vertically-averaged disk (ring), and neglect the returning stream $\dot M_{\rm fb}$ after the circularization, so we only calculate the 1D evolution.

\item We consider the form of advective cooling term is Equation (\ref{Qadv}). However, it is important only if the accretion rate is super-Eddington \citep{Shen_Evolution_2014}.

\item We use the $\alpha_{\rm g}$-viscosity ansatz, i.e., Equation (\ref{viscosity}), to avoid the thermal instability \citep{Lightman_Black_1974}. Moreover, because the fitting results of TDEs in \cite{Van_Velzen_Late_2019} give the high viscosity, i.e., $\alpha > 0.1$, we set $\alpha = 1$ in the following calculations.

\end{itemize}

\begin{figure}
\centering
 \includegraphics[scale=0.5]{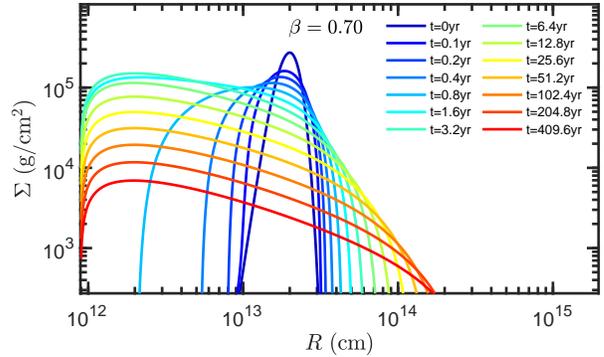}
\caption{Evolution of the disk surface density for the case of $\beta = 0.7$. The colors of the lines represent the time evolution.}
\label{ring_to_disk_sr}
\end{figure}

We show the results of the evolution of the surface density in Figure \ref{ring_to_disk_sr}. We can see the time-scales of the ring-to-disk phase are consistent with the analytical calculation in section \ref{analytic}. Even though the opacity in the numerical calculation is not a constant, the time-scale $t_0$,  i.e., Equation (\ref{t0}) depends weakly on $\kappa$, so $t_0$ is a good approximation of the time-scale of ring-to-disk phase.

\begin{figure}
\centering
 \includegraphics[scale=0.5]{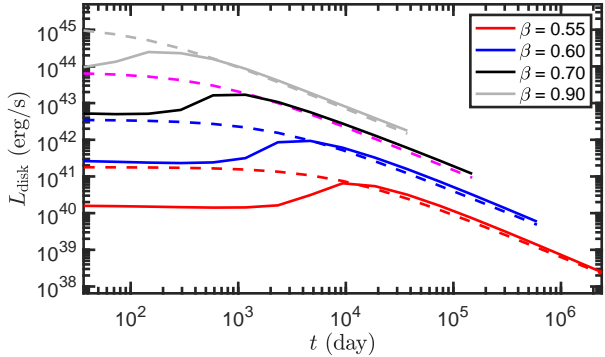}
\caption{Disk bolometric luminosity history for different $\beta$ calculated by the numerical method in section \ref{numerical}. The dashed lines are the analytical results in section \ref{analytic}.}
\label{Lbol}
\end{figure}

The emergent energy flux is given by 
\begin{equation}
F_{\rm vis} = \frac12 Q^-_{\rm rad} = \sigma T_{\rm eff}^4,
\label{Fvis}
\end{equation}
where $\sigma$ and $T_{\rm eff}$ are the Stefan-Boltzmann constant and local effective temperature, respectively. The factor $1/2$ comes form the two side of the disk (ring). The bolometric luminosity is given by 
\begin{equation}
L_{\rm disk} = 2 \times \int^{R_{\rm out}}_{R_{\rm in}} 2\pi R \sigma T_{\rm eff}^4\ dR,
\label{Lb}
\end{equation}
and is shown in Figure \ref{Lbol}. After the formation of accretion disk, the numerical results approach the self-similar ones. However, the analytical results overestimate the luminosity in the ring-to-disk phase.

For an observer at distance $D$ whose viewing angle is $i$, with the blackbody assumption the flux at wavelength $\lambda$ is given by
\begin{equation}
F_{\lambda}=\frac{2\pi\cos{i}}{D^2} \int^{R_{\rm out}}_{R_{\rm in}} B_{\lambda}(T_{\rm eff})R\ dR.
\label{vFv}
\end{equation}
where the Planck function $B_{\lambda}$ is
\begin{equation}
B_{\lambda}=\frac{2h c^2}{\lambda^5 \left(\ue^{hc/(\lambda k_{\rm b} T_{\rm eff})}-1\right)},
\end{equation}
and $h$ is the Planck constant. 

\begin{figure}
\centering
 \includegraphics[scale=0.5]{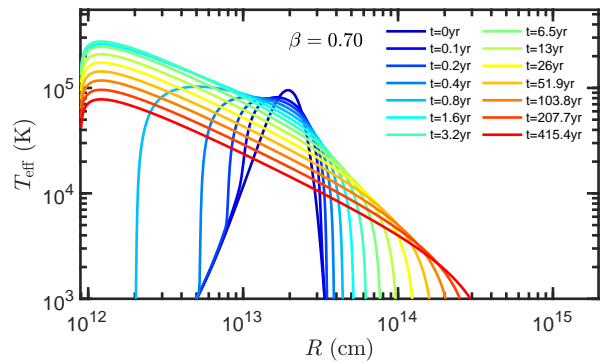}
\caption{Evolution of the local effective temperature distribution on the disk for $\beta = 0.7$.}
\label{local_Teff}
\end{figure}

\begin{figure}
\centering
 \includegraphics[scale=0.5]{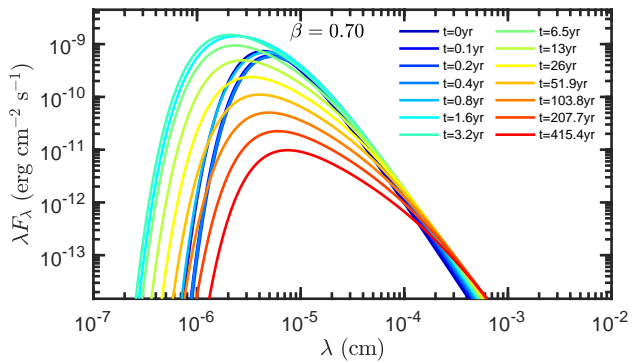}
\caption{Evolution of the disk emission spectrum for $\beta = 0.7$. It is calculated by Equation (\ref{vFv}) assuming viewing angle is $i=0$ and the distance $D = 10\ {\rm Mpc}$. The colors of the lines represent the time evolution. }
\label{ring_to_disk_vFv}
\end{figure}

\begin{figure}
\centering
 \includegraphics[scale=0.5]{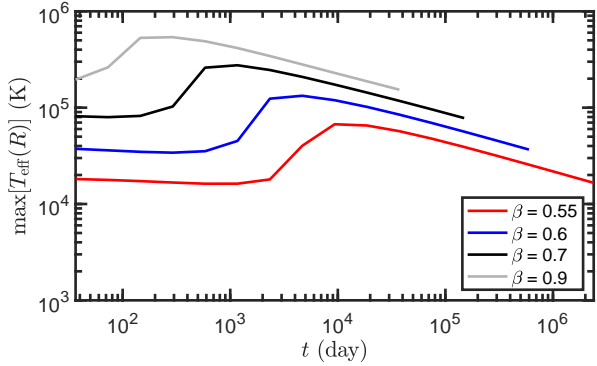}
\caption{Evolution of the peak local effective temperature of the disk for different $\beta$.}
\label{Teff_max}
\end{figure}

The evolution of local effective temperatures and that of the spectrum are shown in Figure \ref{local_Teff} and \ref{ring_to_disk_vFv}, respectively. We use the peak local effective temperature $\max [T_{\rm eff}(R)]$ to represent the observed effective temperature of the whole disk surface, and plot its evolution in Figure \ref{Teff_max}.

\section{Overall light curve and temperature evolution}
\label{temperature_evolution}

\subsection{During Circularization}
\label{circularization_radiation}
During the circularization process, we assume most of the radiation comes from the shocked region, and the shocked matter-radiation mixture is thermalized. Here we estimate the effective temperature by the single blackbody assumption
\begin{equation}
T_{\rm eff} \simeq [L_{\rm c}/(\sigma S)]^{1/4},
\end{equation}
and assume the viewing angle is face-on. We determine the radiative area $S$ by considering the diffusion time-scale as below.

When the radiative diffusion is efficient, i.e., $t_{\rm diff} \lesssim t_{\rm s}$, the initial radiative area $S_0$ can be written as the product of the width and the length of the radiative region: 
\begin{equation}
S_0 \simeq 2w_{\rm s, 0} l_{\rm diff, 0}.
\end{equation}
Here the length can be estimated by $l_{\rm diff, 0} \sim v_{\rm a, 0} t_{\rm diff, 0}$, and $v_{\rm a, 0} \simeq (GM_{\rm h}/a_0)^{1/2}((1-e_0)/(1+e_0))^{1/2}$ is the stream velocity near the opocenter.

We assume the initial expansion of stream is ballistic after the collision, thus the width is $w_{\rm s, 0} \simeq R_0+c_{\rm s, 0} t_{\rm diff, 0}$, where $R_0$ is the initial width of the stream near the collision point and the sound speed after the collision is $c_{\rm s, 0} \simeq 2/3 \Delta \epsilon_0^{1/2}$. For PTDEs, the self-gravity is negligible. Before the collision, the stream width is dominated by the tidal shear, thus in this limit $R_0 \simeq R_*(a_0/R_{\rm p})$ \citep{Kochanek_The_1994,Coughlin_On_2016}.

If $c_{\rm s, 0} t_{\rm diff, 0} \gg R_0$, then it is the former that determines $w_{\rm s, 0}$. We can estimate the initial radiative area as
\begin{equation}
\begin{split}
S_{\rm 0} &\simeq 4 \times 10^{-4}\ \left(\frac{\Delta M}{0.01\ \rm{M_{\odot}}}\right)^2 \times
\\
&\beta^{11} M_6^{-7/6} m_*^{11/3} r_*^{-11/2}\ a_{\rm 0}^2,
\end{split}
\label{area0}
\end{equation}
where we use Equations (\ref{semimajor}), (\ref{eq:delta-eps0}) and (\ref{eq:tdiff}) and assume an initial $h_{\rm s}/w_{\rm s} \simeq 1$.

After the collisions, the orbital energy will be redistributed gradually, causing the stream to extend slightly in the radial direction. Furthermore, the viscous shear in the late stage of the circularization grows, further widening the stream. It is difficult to obtain the detailed evolution of the radiative area by an analytical method. Instead, we parametrize the radiative area evolution in a smooth power-law form
\begin{equation}
S = S_0 \left( \frac{t}{1.5\ t_{\rm fb}}\right)^{\gamma},
\label{area}
\end{equation}
where $1.5\ t_{\rm fb}$ is the time when the first collision occurs. The power law index $\gamma$ is determined by the starting condition of the disk viscous evolution (see below).

\subsection{Circularization Process to Disk Accretion}
\begin{figure}
\centering
 \includegraphics[scale=0.5]{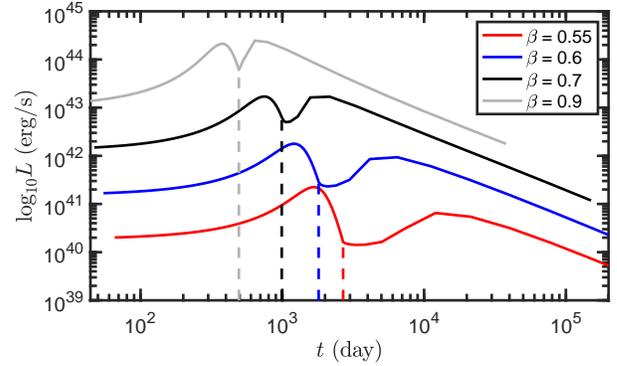}
\caption{Overall bolometric luminosity history for the disruption of a star ($m_*=r_*=1$) by a $10^6\ \rm{M_{\odot}}$ SMBH. The vertical dashed lines represent the time of the transition $t_{\rm c}$ between the circularization stage and the viscous evolution.}
\label{L_whole}
\end{figure}

In order to obtain the whole light curve including the circularization stage and the disk viscous evolution stage, we connect the luminosity curves of the two stages at an intermediate point where these two are equal, as is shown in Figure \ref{L_whole}. After this transition time $t_{\rm c}$, the disk viscous evolution dominates the luminosity.

So far we have assumed the emission spectrum is a single blackbody in the circularization stage and a multi-color blackbody in the disk viscous evolution stage. Thus there are some uncertainties in the transition between the two. In fact, if the initial width of the circularized ring at the beginning of the second stage is small, the spectrum is approximate to a single blackbody as well. Therefore, we expect that the effective temperature should vary smoothly during the transition. Therefore, the radiative area evolution index $\gamma$ during the circularization process can be determined by
\begin{equation}
\gamma = \frac{\log(S_{\rm c}/S_0)}{\log(t_{\rm c}/1.5\ t_{\rm fb})}.
\label{power_law}
\end{equation}
Here $S_{\rm c} = L_{\rm disk}/(\sigma \max[T_{\rm eff}(R)]^4)$ is the effective radiative area at the transition time $t_{\rm c}$ when the luminosity of the viscous evolution becomes dominating the luminosity. 

\begin{figure}
\centering
 \includegraphics[scale=0.5]{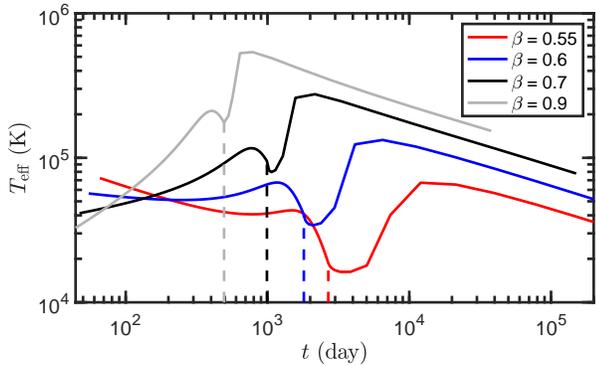}
\caption{Overall effective temperature history. Parameters are same as in Figure \ref{L_whole}. The effective temperature is defined as a single blackbody temperature in the circularization stage, and as the peak local effective temperature of the multi-color blackbody spectrum of the disk in the viscous evolution.}
\label{Teff_whole}
\end{figure}

With these, the overall evolution of the effective temperature for the entire PTDE can be calculated and is plotted in Figure \ref{Teff_whole}.

\section{BH mass and stellar dependence}
\label{dependence}
The circularization time-scale and the peak luminosity in the circularization, i.e., Equations (\ref{circularization}) and (\ref{eq:Lcp}), depend on the BH mass and the stellar properties. Because the effect of general relativity is stronger with higher BH masses, the circularization time-scale is shorter and the luminosity is higher, thus the circularization stage is much easier to be detected. For comparison purpose, here we study the PTDEs with a $10^7 \ \rm{M_{\odot}}$ SMBH.

\begin{figure}
\centering
 \includegraphics[scale=0.5]{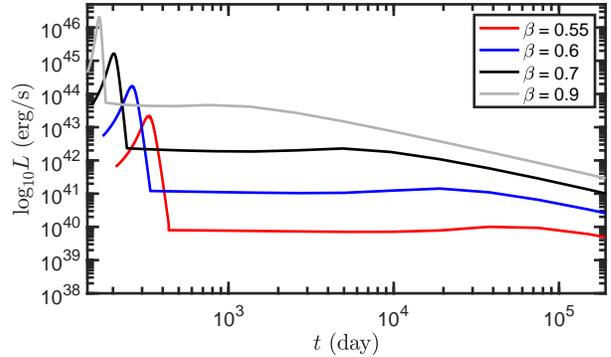}
\caption{Overall bolometric luminosity history for PTDEs with $10^7\ \rm{M_{\odot}}$ SMBH.}
\label{L_whole_m7}
\end{figure}

\begin{figure}
\centering
 \includegraphics[scale=0.5]{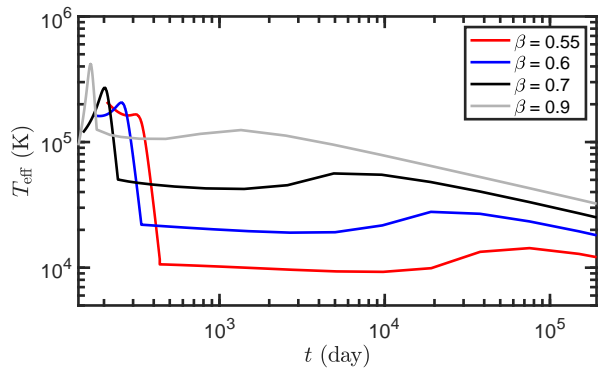}
\caption{Overall effective temperature history for PTDEs with $10^7\ \rm{M_{\odot}}$ SMBH.}
\label{Teff_whole_m7}
\end{figure}

We plot the light curve and the effective temperature evolution for this case in Figures \ref{L_whole_m7} and \ref{Teff_whole_m7}. Compared with the case of a $10^6\ M_{\odot}$ SMBH, the circularization time-scale is shorter, and the luminosity of the circularization stage is higher. On the other hand, the viscous time-scale is longer and the viscous luminosity is lower. The ratio between the luminosities in the circularization and in the viscous evolution is 
\begin{equation}
\begin{split}
\frac{L_{\rm p}}{L_{\rm disk, p}} &\simeq 30\ \alpha^{-4/3} \left(\frac{\Delta M}{0.01\ \rm{M_{\odot}}} \right)^{-2/3} \times
\\
&\beta^{13/6} M_6^{28/9} m_*^{13/18} r_*^{-13/6},
\end{split}
\end{equation}
which is very sensitive to the BH mass. The luminosity in the viscous evolution is very weak for a $10^7\ M_{\odot}$ SMBH.

Most of the tidally disrupted stars come from the lower end of the stellar mass function \citep{Stone_Rates_2016,Kochanek_Tidal_2016}. Here we also calculate the case of a smaller main-sequence star with mass $m_*=0.2$ using the mass-radius relation $r_*=m_*^{0.89}$ \citep{Torres_Accurate_2010}.

\begin{figure}
\centering
 \includegraphics[scale=0.5]{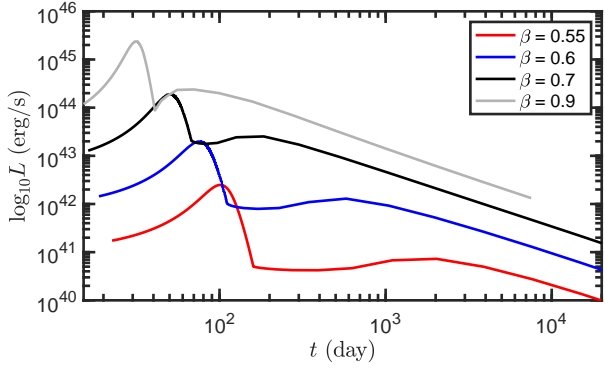}
\caption{Overall bolometric luminosity history for PTDEs of a low-mass star ($m_*= 0.2$).}
\label{L_whole_02}
\end{figure}

\begin{figure}
\centering
 \includegraphics[scale=0.5]{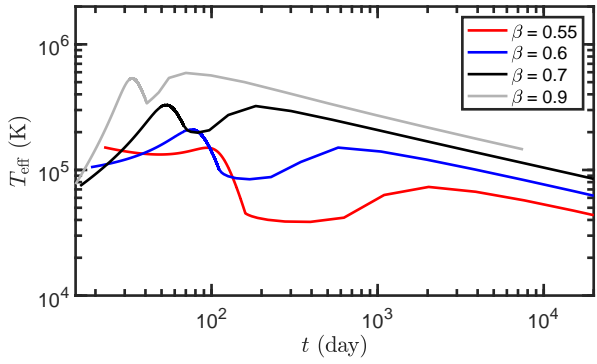}
\caption{Overall effective temperature history for PTDEs of a low-mass star ($m_*= 0.2$).}
\label{Teff_whole_02}
\end{figure}

We plot the light curve and the effective temperature evolution in Figures \ref{L_whole_02} and \ref{Teff_whole_02}, respectively. The luminosity is little higher than that of $m_*= 1$ (see Equation (\ref{eq:Lcp})), and the circularization time-scale and viscous time-scale are shorter. The evolution of temperature is similar to that of $m_*= 1$.

The luminosity and the temperature of borderline FTDE with $\beta = 0.9$ we plot here are just for comparison. Since $t_{\rm diff}/t_{\rm s} \propto m_*^{-0.56} M_{\rm h}^{-7/6}$, it is insensitive to the stellar mass and is lower for the larger BHs, and $t_{\rm diff}/t_{\rm s} \gtrsim 1$ for FTDEs (see the Figure \ref{diffusion}). Thus the photon diffusion of FTDE is inefficient, the stream will expand intensively after the intersections of the debris streams, and an elliptical disk might be formed \citep{Shiokawa_General_2015,Piran_Disk_2015,Liu_Elliptical_2020}. Furthermore, the accretion process is super-Eddington for $\beta \gtrsim 0.9$, thus it will produce disk wind and affect the light curve and the spectrum. These features are not accounted for in our calculation.

\section{Event rate and Detection rate}
\label{event_detection_rate}
\subsection{Event rate}
The parameter space for PTDEs and FTDEs are $0.5 < \beta < \beta_{\rm d}$ and $\beta_{\rm d} < \beta < \beta_{\rm max}$, respectively. The exact value of $\beta$ for the partial disruption onset is $\beta > 0.5$ \citep{Ryu_Tidal1_2020}.The upper limit of the penetration factor for disruption $\beta_{\rm max}$ is given by $\beta_{\rm max} \simeq R_{\rm T}/R_{\rm S}$. If $\beta > \beta_{\rm max}$, SMBH will directly swallow the whole star instead of tidally disrupting it \citep{Kesden_Tidal_2012}. Notice that if $\beta_{\rm max}<\beta_{\rm d}$, only PTDEs can occur in that case.

We can estimate the event rate of TDEs by the loss-cone dynamics \citep{Merritt_Loss_2013}. Occasionally a star will be scattered into a highly eccentric orbit with pericenter radius $R_{\rm p} \lesssim R_{\rm lc} \equiv \max[R_{\rm S}, R_{\rm d}]$, and the star will be captured or fully disrupted by the SMBH. Here $R_{\rm d} = R_{\rm T}/\beta_{\rm d}$ is the full disruption radius. The rate of the stars entering $R_{\rm lc}$ (hereafter as $\dot N_{\rm lc}$) depends on the realistic stellar density profile in the nucleus of each galaxy. In \cite{Stone_Rates_2016}, they use an early-type galaxy sample consisting of 144 galaxies to calculate the TDE rate. Here we use their fitting result (Eq. 27 in \cite{Stone_Rates_2016}), i.e.,
\begin{equation}
\label{eq:erate}
\dot N_{\rm lc} = \dot N_0 \left(\frac{M_{\rm h}}{10^8\ M_{\odot}}\right)^B,
\end{equation}
with $\dot N_0 = 2.9 \times 10^{-5}\ \rm{yr^{-1} gal^{-1}}$ and $B = -0.404$, to estimate the stellar loss rate in a galaxy.

Using the loss rate $\dot N_{\rm lc}$, the fraction function of SMBHs $\phi(M_{\rm h})$, the fraction function of penetration factor $f_{\rm TDE}$, and the initial mass function (IMF) $\chi_{\rm Kro}$ in Appendix \ref{appendix}, one can calculate the differential volumetric event rates of PTDEs and FTDEs with respect to the SMBH mass by
\begin{equation}
\label{eq:volrate}
\frac{d \dot N_{\rm TDE}}{dM_{\rm h}} = \int^1_{m_{*, {\rm min}}} \int_{\beta} \dot N_{\rm lc} \phi(M_{\rm h}) \xi_{\rm Kro}(m_*) f_{\rm TDE}\ dm_* d\beta,
\end{equation}
where $m_{*, {\rm min}}$ is the lower limit for TDE, it is given by Eq. (\ref{eq:mstarmin}).

\begin{figure}
\centering
\includegraphics[scale=0.5]{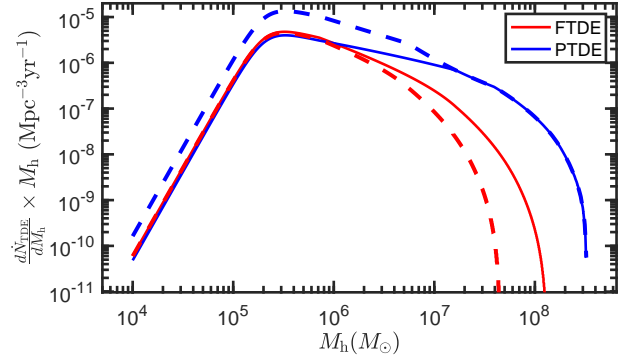}
\caption{Volumetric event rate of TDEs versus SMBH mass $M_{\rm h}$. The solid and dashed lines represent $\gamma = 5/3$ and $\gamma=4/3$ polytropic indices of stars, respectively.}
\label{volrate}
\end{figure}

The volumetric event rates of TDEs with respect to $M_{\rm h}$ is shown in Fig. \ref{volrate}. The structure of polytropic stars with $\gamma = 4/3$ are denser than that with $\gamma = 5/3$, therefore it needs to be much closer to the SMBH for full disruptions, thus reducing the probability for FTDEs.

The event rates of PTDE and FTDE are similar for the smaller SMBHs $M_{\rm h} \le 10^6\ M_{\odot}$. However, the event rate of PTDEs becomes dominant for the larger SMBHs. There are two reasons: First, the radius for the full disruption become closer to the SMBH as the SMBH mass increases, when $M_{\rm h} > 10^8\ M_{\odot}$ the stars can not be fully disrupted, then only PTDEs can occur. Second, most of the stars are in the diffusion limit for the lager SMBHs (see Eq. (\ref{eq:fpin})), which will suppress the FTDEs.

Overall, the chance of PTDEs is very promising. The event rate of PTDEs is greater than that of FTDEs, even for the larger SMBHs. Notice that \cite{Ryu_Tidal1_2020} consider the realistic stellar structure by using MESA, and find that for low-mass stars the chance of PTDEs is approximately equal to that of FTDEs, but for high-mass star, the likelihood of PTDEs is 4 times higher than that of FTDEs. However, they consider only the full loss-cone regime and neglect the upper limit of penetration $\beta_{\rm max}$ for FTDEs. Therefore, their estimate of the PTDE fraction is lower than ours.

\subsection{Detection rate of PTDEs}
The fallback mass of PTDE is less than that of FTDE, so PTDE is seemingly dimmer than FTDE. However, as we discuss above, PTDEs prefer heavier SMBHs, thus most of them are bright and detectable. The detection rate depends on the physical processes that contribute to the emission of TDEs \citep{Stone_Rates_2016}. Here we calculate the detection rate of PTDEs using our model.

The limiting detection distance for a PTDE is
\begin{equation}
\label{dlim}
d_{\rm lim} = \left[\frac{L_{\nu}}{4\pi f_{\nu}}\right]^{1/2},
\end{equation}
where the spectral flux density limit $f_{\nu}$ is given by $m_{\rm R} \simeq -2.5\ \lg(f_{\nu}/3631 \rm{J_y})$ and the R-band limiting magnitude $m_R \simeq 20.5$ for ZTF. The monochromatic luminosity of the source is $L_{\nu} \simeq \pi B_{\nu}(T_{\rm eff}) S$, where $B_{\nu}$ is the Planck function. The effective temperature $T_{\rm eff} \sim (L_{\rm p}/(\sigma S_0))^{1/4}$ and radiative area $S \sim S_0$ are given in the model, i.e., Eq. (\ref{eq:Lcp}) and (\ref{area0}), thus $d_{\rm lim}$ is the function of $\beta$, $M_{\rm h}$ and $m_*$.

The detection rate of PTDEs is
\begin{equation}
\label{eq:detection}
\begin{split}
D_{\rm p} \simeq \int dM_{\rm h} \int_{m_{\rm *, min}}^1 dm_* &\int_{0.5}^{\beta_{\rm lc}} \dot N_{\rm lc}(M_{\rm h})\phi(M_{\rm h})\chi_{\rm Kro}(m_*) \\
& \times f_{\rm TDE} \frac43 \pi d_{\rm lim}^3\ d\beta.
\end{split}
\end{equation}
Here $\beta_{\rm lc} \equiv R_{\rm T}/R_{\rm lc}$. The integral upper limit of the SMBH mass is given by $R_{\rm T}/0.5 = R_{\rm S}$, i.e., $M_{\rm h,max} = (R_{\odot}c^2/(GM_{\odot}^{1/3}))^{3/2}$.

It gives $D_{\rm p} \sim 5 \times 10^2\ {\rm yr^{-1}}$ for $\gamma = 5/3$, and $D_{\rm p} \sim 10^2\ {\rm yr^{-1}}$ for $\gamma = 4/3$. Taking the field of view of ZTF into consideration ($\sim 0.1$ of the whole sky), the ZTF detection rate of PTDEs is about dozens per year.

\section{Conclusion and Discussion}
\label{conclusion}
In this paper, we consider that PTDEs may not produce outflow or wind during the circularization due to the efficient radiative diffusion in the stream. Because PTDEs have less fallback mass than FTDEs, photons diffuse out more efficiently. Hence they provide a clean environment to study the circularization process and the disk formation.

We calculate the light curves of PTDEs considering the earlier circularization process and the later disk viscous evolution. During the circularization process, the radiation comes directly from the shocked stream. After the circularization, the ring at the circularization radius evolves by the viscous shear, and eventually settles in a self-similar and sub-Eddington accretion phase.

There are two peaks in the light curve of a PTDE. The first one corresponds to the circularization process, the second one to the formation of the accretion disk. The times of the peaks are $t_{\rm cir} \sim 10^2 - 10^3\ {\rm day}$ and $t_0 \sim 10^3 - 10^4\ {\rm day}$, respectively. Formulae for the time-scales of both phases are provided. The ratio between them is
\begin{equation}
\begin{split}
\frac{t_{\rm cir}}{t_0} &\simeq 0.07\ \alpha^{4/3} \kappa^{1/3} \left(\frac{\Delta M}{0.01\ M_{\odot}}\right)^{2/3} \times
\\
&\beta^{-5/3} M_6^{-41/18} m_*^{-5/9} r_*^{7/6}.
\end{split}
\end{equation}
For most of the PTDEs with $M_{\rm h} \gtrsim 10^6\ M_{\odot}$ and $m_* \gtrsim 0.1$, the circularization time-scale is shorter than the viscous time-scale. Therefore, we conclude that accretion disk forms after the circularization, and we can see the double peaks in the light curve of PTDE.

Increasing either the BH mass, density of the disrupted star, or the penetration factor can enhance the self-crossing shock, because all these bring the pericenter radius closer (in units of the Schwarzschild radius) to the BH; therefore, the luminosity increases and the time-scale is shorter for the circularization stage. In the viscous evolution stage, the viscous time-scale is shorter and the luminosity is higher for a smaller BH mass, a smaller star or a larger penetration factor.  

Based on the single blackbody assumption in the circularization stage, we calculate the effective temperature which follows the light curve to rise and drop. After that, as the circularized ring evolves to an accretion disk, the effective temperature rises until the disk has formed. Eventually, both the light curve and the effective temperature decay in power laws with time, following the self-similar solution of disk evolution. Overall, the effective temperatures are $\sim 10^4 - 10^6\ {\rm K}$ and exhibit weak dependence on the BH mass and the star, so the spectra peak in UV.

\subsection{Viscous Effects in the Circularization Stage}
\label{subsec:vis_cir}
In the calculation of the circularization process, we neglect the viscous effects. Here we explore the importance of viscous effects in the circularization stage. There are two main effects of the viscous shear. One is that the viscous shear can heat up the stream and increase the luminosity. Furthermore, the viscous shear can redistribute the angular momentum of the debris stream and cause some parts of the stream to be closer to the SMBH. Then it may further enhance the dissipation caused by the viscous shear and the self-crossing, and speeds up the formation of the disk (elliptical or circular disk).

\cite{Svirski_Elliptical_2017} \citep[also see][]{Bonnerot_Long_2017} calculate the viscous effects induced by the magnetic stress, which originates from the exponential growth of the magneto-rotational instability (MRI) \citep{Balbus_A_1991}. In order to be consistent with the prescription of viscosity adopted in the disk stage, we also adopt the $\alpha_{\rm g}$-viscosity (Equation (\ref{viscosity})) to parametrize the viscous shear in the circularization stage. 

One can find that the prescription of $\alpha_{\rm g}$-viscosity is equivalent to that of magnetic stress used in \cite{Svirski_Elliptical_2017} if we let $\alpha \simeq \alpha_{\rm mag} (v_{\rm A}/c_{\rm s})^2$. Here $\alpha_{\rm mag}$ and $v_{\rm A}$ are the ratio of the  $\hat n-\hat t$ magnetic stress component to the total magnetic stress and the Alfv\'{e}n velocity, respectively. 

The redistribution of the specific angular momentum caused by the viscous shear can be estimated by $dj/dt \simeq \nu \Omega$. The total change of the specific angular momentum during the circularization process can be written as $\Delta j \simeq \nu \Omega t_{\rm cir}$, thus
\begin{equation}
\begin{split}
\frac{\Delta j}{j} \sim &\alpha t_{\rm cir} \frac{k_{\rm b} T_{\rm c}}{\mu m_{\rm p}j}
\\
&\sim 0.01\ \alpha \left(\frac{T_{\rm c}}{10^6\ {\rm K}}\right) \beta^{-7/2} M_6^{-8/3} m_*^{-7/6} r_*^3,
\end{split}
\end{equation}
where the specific angular momentum is $j \simeq (GM_{\rm h}R_{\rm c})^{1/2}$. For PTDEs with $M_{\rm h} \gtrsim 10^6\ M_{\odot}$, the viscous shear has negligible effect on the extension of stream during the circularization process. Therefore the structure of stream will keep thin until its orbit circularizes and settles into a ring.

Another viscous effect is that it can heat up the stream and increase the luminosity. We can estimate the viscous luminosity in the circularization process by assuming the viscous heating rate equals to the radiative cooling rate, i.e., $L_{\rm c, vis} \simeq \int \nu \Omega^2 \Sigma\ dS$. Most of the viscous heating come from the pericenter where only a small part of the stream mass locates at, thus the viscous luminosity is very small, i.e., $L_{\rm c, vis} \ll (\nu \Omega^2)_{\rm p} \Delta M \simeq L_{\rm disk, 0}$. Here the subscript $p$ denotes the value at the pericenter radius, and $L_{\rm disk, 0}$ is the luminosity of the ring after the circularization process.

Therefore, it is reasonable to neglect the viscous effects in the circularization process. In the late time of the circularization, the viscous effects become important and the luminosity is dominated by the viscous shear.

In the transition between the circularization stage and the viscous evolution stage, we artificially connect the light curve of these two stages due to the fact that the luminosity induced by the viscous shear is small in the circularization stage. In fact the transition should be smooth if we take the viscous heating into account in the circularization stage.

We consider that in the circularization stage most of the radiation comes from the shock-heated debris stream. As the material cools down, the ions will recombine. Assuming most of the material is hydrogen, the total recombination energy is $E_{\rm re}\simeq N \times 13.6\ {\rm eV} \simeq 3 \times 10^{44}\ (\Delta M/0.01\ M_{\odot})\  {\rm erg}$, where $N$ is the number of hydrogen ions. The recombination luminosity is $L_{\rm re} \simeq E_{\rm re}/t_{\rm fb} \simeq 10^{38}\ {\rm erg/s}$, which is negligible. Recombination will probably promote the chemical reactions in the stream, so that dust clumps form \citep{Kochanek_The_1994}. However, these dust will be evaporated in later shocks.

\subsection{Observational Prospect}
\label{subsec:obs}
We calculate the detection rate of PTDEs through loss-cone dynamic. For ZTF, the detection rate is about dozens per year, it is very promising. We encourage the search of them by optical/UV or soft X-ray telescopes. For some PTDEs that might have already been discovered in the past data, our work would be useful to identify them.

Recently, \cite{Gomez_The_2020} report a TDE candidate AT 2018hyz. They use the Modular Open-Source Fitter for Transients (MOSFIT) to model the light curves, and conclude that it is a PTDE. The MOSFIT model assumes a rapid circularization of the stream and that the evolution of luminosity traces the mass fallback rate, which may not be the case for PTDEs, as we have shown. The double peaks in the light curve of AT 2018hyz are consistent with what we predict for a PTDE, which is the feature of a two-stage evolution.

However, the blackbody spectral fitting of AT 2018hyz gives a large photosphere $\sim 10^{15}\ {\rm cm}$, which is not consistent with our model. There are some uncertainty in the spectrum fitting, e.g., the galaxy extinction and prior spectrum assumption. Furthermore, in the circularization process, the spectrum might deviate from the blackbody spectrum we assume here. More details of the radiative dynamical evolution of the circularization need to be understood.

Recently, \cite{Frederick_A_2020} report five transient events found by ZTF from active galactic nucleus (AGNs). One of them, ZTF19aaiqmgl, shows two peaks in the light curve, and only the second peak has X-ray detection. The second peak might correspond to the disk formation. However, in an AGN, the debris from the disrupted star will collide with the pre-existing accretion disk \citep{Chan_Tidal_2019}, and the shocks will heat up the gas. For PTDEs, the lighter streams might directly merge with the disk. The details of PTDEs from AGNs are still unclear.

Furthermore, \cite{Payne_14ko_2021} report a repeated PTDE candidate ASASSN-14ko, which is located within an AGN. The star will be tidally stripped by the BH during each encounter near the pericenter, if the star is in an elliptical orbit. The pre-existing accretion disk can produce periodic bright flares by accreting these stripped mass. Thus the debris might not experience a long-term circularization and disk formation process, but how the debris interact with the pre-existing disk is unclear.

\section{acknowledgments}
We thank the referee for helpful comments and suggestions. This work is supported by the National Natural Science Foundation of China (12073091), Guangdong Basic and Applied Basic Research Foundation (2019A1515011119) and Guangdong Major Project of Basic and Applied Basic Research (2019B030302001).

\begin{appendix}
\section{Functions in the calculation of event rate}
\label{appendix}
In order to calculate the volumetric event rate of TDEs, one needs to consider the fraction of SMBHs with mass $M_{\rm h}$, i.e., $\phi(M_{\rm h})$. We only consider those TDEs occurring near the SMBHs in galaxy nucleus, thus $\phi(M_{\rm h})$ actually is the fraction of galaxies which have SMBHs with mass $M_{\rm h}$.

In \cite{Stone_Rates_2016}, they calculate $\phi(M_{\rm h})$ using the Schechter function \citep[galaxy luminosity function,][]{Schechter_An_1976}, the scaling relations of BH masses and host galaxy properties \citep{McConnell_Revisiting_2013}, and the occupation fraction of SMBHs \citep{Miller_Xray_2015}. We rewrite the fraction function of SMBHs here
\begin{equation}
\label{eq:schefunm}
\begin{split}
\phi(M_{\rm h}) dM_{\rm h} &= 3.53 \phi_* f_{\rm occ} M_6^{-1.07} \\
& \times \exp \left(-0.025M_6^{0.709}\right) dM_6\end{split}
\end{equation}
where $\phi_*= 4.9 \times 10^{-3} h_7^{3}\ {\rm Mpc^{-3}}$, $f_{\rm occ}(M_{\rm h})$ is the occupation fraction of SMBHs, and we take the normalized Hubble constant $h_7 = 1$. 

The occupation fraction of SMBHs $f_{\rm occ}$ is the probability that a galaxy harbors a SMBH, which is given by \citep{Miller_Xray_2015}
\begin{equation}
\label{eq:focc}
\begin{small}
f_{\rm occ} = \begin{cases}
0.5+0.5&\tanh \left(\ln\left(\frac{M_{\rm bul}}{M_{\rm c}}\right) \times
2.5^{8.9-\log_{10}\left(\frac{M_{\rm c}}{M_{\odot}}\right)}\right), \\
&{M_{\rm bul} < 10^{10} M_{\odot}} \\
1,&{M_{\rm bul} > 10^{10} M_{\odot}},
\end{cases}
\end{small}
\end{equation}
where $M_{\rm bul}$ is the bulge mass, which we relate to the SMBH mass using the $M_{\rm bul}-M_{\rm h}$ relation from \cite{McConnell_Revisiting_2013}, i.e.,
\begin{equation}
\label{Mbulge}
\log_{10}(M_{\rm h})=8.46+1.05\log_{10}(M_{\rm bul}/10^{11}\ M_{\odot}).
\end{equation}
The parameter $M_{\rm c}$ is the approximate mass below which the occupation fraction turns over. It should be less than $\sim 10^{8.5}\ M_{\odot}$ \citep{Stone_Rates_2016}. Here we assume $M_{\rm c} \simeq 10^8\ M_{\odot}$. The exact value of $M_{\rm c}$ only affects the occupation fraction for the galaxies with smaller SMBHs. As we shall see, most of the observable PTDEs occur near the larger SMBHs, therefore it changes little the detection rate of PTDEs. 

Furthermore, only those stars with low density can be disrupted by the SMBH. That is because their disruption radius are outside the horizon. FTDEs and PTDEs require $R_{\rm d} \gtrsim R_{\rm S}$ and $R_{\rm T}/0.5 \gtrsim R_{\rm S}$, respectively. Using the stellar mass-radius relation for the lower main sequence $r_* \propto m_*^{0.89}$ \citep{Torres_Accurate_2010}, one can obtain the lower limit of stellar mass for disruption is
\begin{equation}
\label{eq:mstarmin}
m_{*,min} = \begin{cases}
0.85\ \beta_{\rm d}^{1.8} \left(\frac{M_{\rm h}}{10^8 M_{\odot}}\right)^{1.2},&\quad{\rm FTDEs} \\
0.25\ \left(\frac{M_{\rm h}}{10^8 M_{\odot}}\right)^{1.2},&\quad{\rm PTDEs}.
\end{cases}
\end{equation}

The stars have large orbital period would diffuse across the loss-cone by gravitational encounters in a single orbit, i.e., the so-called full loss-cone regime or pinhole limit. And the stars near the SMBH, have short orbital period, will diffuse into the loss-cone over many orbits, and thus hardly penetrate beyond the loss-cone boundary, i.e., the so-called empty loss-cone regime or diffusion limit.

Unlike the calculation in \cite{Stone_Rates_2016}, we consider that the stars in the diffusion limit and in the pinhole limit have different fates. In the diffusion limit, most of the stars experience one or more partial disruptions as they approaching the loss cone. After the partial disruption, their remnant cores probably return to be disrupted again or escape as the so-called ''turbovelocity'' stars \citep{Manukian_Turbovelocity_2013,Ryu_Tidal3_2020}, anyway, that will result in the strong suppression of FTDEs in the diffusion limit.

And in the pinhole limit, the velocity directions of stars are randomly distributed, then the fraction function with $\beta$ is $f_{\rm TDE} \propto \beta^{-2}$. Therefore, for the FTDE ($\beta_{\rm d} < \beta < \beta_{\rm max}$), we assume only the stars in pinhole limit can be fully disrupted, thus $f_{\rm TDE} \propto f_{\rm pin}\beta^{-2}$. Here the pinhole fraction $f_{\rm pin}$ is the fraction of the stars whose orbits are in the pinhole limit near the SMBH. It can be estimate by the fitting formula \cite[i.e., Eq. (29) in][]{Stone_Rates_2016}
\begin{equation}
\label{eq:fpin}
f_{\rm pin} = 0.22 \left(\frac{M_{\rm h}}{10^8 M_{\odot}}\right)^{-0.307},
\end{equation}
which should satisfy $f_{\rm pin} < 1$. 

For the PTDE ($0.5 < \beta < \beta_{\rm lc}$), we assume $f_{\rm TDE}$ contains the contributions of stars in both diffusion limit and pinhole limit. In the pinhole limit, $f_{\rm TDE} \propto f_{\rm pin}\beta^{-2}$. In the diffusion limit, we assume the stars outside the loss-cone ($R_{\rm lc}$) are in the quasi-steady state, then the distribution of angular momenta of stars can be obtained by the steady-state solution, i.e., $\xi_{\rm diff}(\mathcal{R}) \propto \ln(\mathcal{R})$ \citep{Merritt_Loss_2013}, Here $\mathcal{R} \equiv j^2/j_{\rm lc}^2$ with $j_{\rm lc} \simeq (2GM_{\rm h}R_{\rm lc})^{1/2}$. It satisfies $ \xi_{\rm diff}(\mathcal{R})\ d\mathcal{R} =  \xi_{\rm diff}(\beta)\ d\beta$ and
\begin{equation}
\int^{\beta_{\rm lc}}_{0.5} \xi_{\rm diff}(\beta)\ d\beta  = 1.
\end{equation}
Thus we have
\begin{equation}
\label{eq:xidff}
\xi_{\rm diff}(\beta) = \frac{0.5 \ln{(\mathcal{R})}}{\ln{(\beta_{\rm lc}/0.5)}+0.5/\beta_{\rm lc}-1}\beta^{-2}.
\end{equation}

Then we obtain the fraction function with $\beta$ in TDE is
\begin{equation}
f_{\rm TDE} = \begin{cases}
(1-f_{\rm pin})\xi_{\rm diff}(\beta)+f_{\rm pin}\frac{\beta^{-2}}{1/\beta_{\rm lc}},&{\rm PTDEs} \\
f_{\rm pin}\frac{\beta^{-2}}{1/\beta_{\rm d}},&{\rm FTDEs}.
\end{cases}
\end{equation}

Furthermore, because the TDE rate depends on the present-day mass function of stars, we need to take the IMF into account. We adopt the Kroupa IMF \citep{Kroupa_On_2001}, i.e.,
\begin{equation}
\label{eq:kroupa}
\chi_{\rm Kro} = \begin{cases}
0.28m_*^{-1.3},&\quad{0.08 < m_* < 0.5} \\
0.14m_*^{-2.3},&\quad{0.5 < m_* < 1} \\
0,&\quad{\rm {otherwise}},
\end{cases}
\end{equation}
where the upper truncation $m_*=1$ was chosen to approximate an old stellar population. It satisfies $\int \chi_{\rm Kro}\ dm_* = 1$.

\end{appendix}

\bibliography{cited}

\end{CJK*}
\end{document}